\documentclass[onecollarge]{svjour2}       % onecolumn "king-size"
\smartqed  % flush right qed marks, e.g. at end of proof
\usepackage{amssymb}
\usepackage{amsmath}
\usepackage[dvips]{graphicx}
\usepackage{bm}
\usepackage{amsbsy}
 \journalname{Continuum Mechanics and Thermodynamics}
\begin{document}

\title{On the M\"uller   paradox for thermal-incompressible media%\thanks{Grants or other notes
%about the article that should go on the front page should be
%placed here. General acknowledgments should be placed at the end of the article.}
}

%\titlerunning{}        % if too long for running head

\author{H. Gouin         \and
        A. Muracchini \and
        T. Ruggeri\footnote{corresponding author} %etc.
}

%\authorrunning{Short form of author list} % if too long for running head

\institute{Henri  Gouin \at
              C.N.R.S. U.M.R. 6181 \&
 University of Aix-Marseille,   Case 322, Av. Escadrille
 Normandie-Niemen, 13397 Marseille Cedex 20 France.\\
              \email{henri.gouin@univ-cezanne.fr}           %  \\
%             \emph{Present address:} of F. Author  %  if needed
           \and
           Augusto Muracchini and Tommaso Ruggeri  \at
              Department of Mathematics and Research Center of Applied Mathematics
University of Bologna, Via Saragozza 8, 40123 Bologna, Italy.\\
\email{augusto.muracchini@unibo.it; tommaso.ruggeri@unibo.it}}

%\date{Received: date / Accepted: date}
\date{}
% The correct dates will be entered by the editor

\maketitle

\begin{abstract}
In his  monograph \emph{Thermodynamics}, I. M\"uller proves that for
incompressible media the volume does not change with the temperature. This
M\"uller paradox yields an incompatibility between experimental evidence and
the entropy principle. This result has generated much debate within the
mathematical and thermodynamical communities as to the basis of Boussinesq
approximation in fluid dynamics.\newline  The aim of this paper is to prove
that for an appropriate definition of incompressibility, as a limiting case of
\textit{quasi thermal-incompressible} body, the entropy principle holds for
pressures smaller than a critical pressure value. The main consequence of our
result is the physically obvious one, that for very large pressures, no body
can be perfectly incompressible. The result is first established in the fluid
case.  In the case of hyperelastic  media subject to large deformations the
approach is similar, but with a suitable definition of the pressure associated
with convenient stress tensor decomposition. 
\end{abstract}

\section{Introduction}

It is well known that compressible and incompressible bodies have different
mathematical treatments: for   compressible media the pressure is a constitutive
function, while for incompressible media the pressure comes from a Lagrange
multiplier associated with the constraint of incompressibility. From an
experimental point of view, incompressible medium has no real existence but
can be approximated as limit case of the compressible one. Starting from this
observation much literature has been devoted by using qualitative
analysis and numerical methods  to search solutions of the incompressible case; for example,
the limit of solutions of the compressible fluids are considered as the Mach number tends to
zero under certain assumptions on the initial data (see e.g.   the
isothermal case \cite{kl,klm,LiMa,Des}). When the thermal effects are relevant
the limit is more ambiguous and depends on the particular model of
incompressibility (see e.g. \cite{Bech}). With the aim to have the same
set of equations for compressible and incompressible fluids, M\"{u}ller
\cite{Muller1} and others (see e.g. \cite{Raj,Raj1}) choose the pressure $p$ as
unknown field variable instead of the density $\rho $. More precisely, in the
case of dissipative Navier-Stokes-Fourier fluids, they add to the balance law system of mass,
momentum and energy,
\begin{eqnarray}
&&\frac{\partial \rho }{\partial t}+\frac{\partial \rho v_{i}}{\partial x_{i}%
}=0,   \\
&&    \\
&&\frac{\partial \rho v_{j}}{\partial t}+\frac{\partial }{\partial x_{i}}%
(\rho v_{i}v_{j}-t_{ij})=\rho f_{j},
\label{mixture} \\
&&    \\
&&\frac{\partial \rho \left( \varepsilon +\frac{1}{2}v^{2}\right) }{\partial
t}+\frac{\partial }{\partial x_{i}}\left\{ \rho \left(\varepsilon +\frac{1}{2}%
v^{2}\right)v_{i}-t_{ij}v_{j}+q_{i}\right\} =\rho f_{j}v_{j}+\rho r,
\end{eqnarray}%
the following constitutive equations,
\begin{equation}
\rho \equiv \rho (p,T),\quad \varepsilon \equiv \varepsilon (p,T),
\label{const1}
\end{equation}%
\begin{equation}
\sigma_{ij}=\nu (p,T)\,d_{kk}\delta _{ij}+2\mu (p,T)\,d_{ij},\quad
q_{i}=-\kappa (p,T)\,\partial T/\partial x_i\, ,  \label{const0}
\end{equation}
with
\begin{equation}
\textbf{t}=-p\textbf{1} + \bm{\sigma}\,, \label{fluidtensor}
\end{equation}
where $\textbf{t}\equiv (t_{ij})$ represents the
stress tensor, $\bm{\sigma}\equiv({\sigma}_{ij})$ is the viscous shear
stress, $\textbf{1}\equiv (\delta_{ij})$ the identity tensor and $\delta_{ij}$
the Kronecker symbol. The other variables have the usual meaning:
$\mathbf{v}\equiv (v_{j}),\, \mathbf{q}\equiv (q_{j})$,  $\varepsilon$, $T$
respectively denote the velocity, the heat flux, the internal energy and the
temperature while $\mathbf{f}\equiv (f_{j}), \, r$, are the specific body force
and the heat supply; $\kappa$ is the heat conductivity and the scalars $\nu $
and $\mu $ are the viscosity coefficients. Matrix $\mathbf{d}=\Vert d_{ij}\Vert
$ denotes the symmetric part of $\nabla
\mathbf{v}$ ($d_{kk}=\rm{div}\,\,\mathbf{v}$). Consequently,  system (%
\ref{mixture})-(\ref{const0}) is a closed system for the unknown variables $%
p $, $T$, $\mathbf{v}$. \newline M\"{u}ller defines an incompressible fluid as
a medium for which the constitutive equations (\ref{const1})-(\ref{const0}) are
independent of the pressure, and  in particular:
\begin{equation}
\rho \equiv \rho (T)\,\,,\quad \varepsilon \equiv \varepsilon (T)\,.
\label{indp}
\end{equation}%
Nonetheless, he proves \cite{Muller1} that the only function $\rho (T)$
compatible with the entropy principle is a constant function $\rho =\rho_o$.
Obviously, this result disagrees with experiments showing that the density changes with the temperature (see  e.g.  \cite{Handbook}) and with all
the theoretical results given in particular in the so-called Boussinesq
approximation (see  e.g. \cite{Chandra,brian}). We call this
contradiction the \textit{M\"{u}ller paradox}. \newline In this paper we prove
that for a convenient definition of incompressibility, the entropy principle is
compatible with the fact that volume changes with temperature provided that the
pressure is  smaller than a critical pressure value $p_{cr}$.
These results provide a precise value of the critical pressure under which a fluid can be experimentally
similar to an incompressible one and   permits  to obtain a quantitative
measurement for which the Boussinesq approximation can be considered as valid.
\newline
The paper is organized as follows:\newline In Section 2, the M\"uller paradox
is examined in fluid case.\newline In Section 3, we prove that exists a critical
pressure $p_{cr}$ such that for any $p\ll p_{cr}$ the paradox is solved. Then,   a
numerical evaluation is done allowing  to obtain the
critical pressure value for which water can be considered as  incompressible
liquid.\newline In section 4, the case of hyperelastic media subject to large
deformations is considered; thanks to a stress tensor decomposition carrying
out a suitable definition of the pressure, an approach similar to fluid case
allows to solve the M\"uller paradox. We perform a numerical
evaluation of the critical pressure for pure gum rubber.

\section{The M\"uller paradox for incompressible fluids}

The Gibbs relation
\[
TdS=d \varepsilon-\frac{p}{\rho^{2}}d\rho,
\]
where $S$ is the entropy density, can be rewritten by using the chemical potential
\begin{equation}
\mu= \varepsilon+pV-TS \, ,  \label{pot}
\end{equation}
in the form
\begin{equation}
d\mu= Vdp -SdT\,, \label{dpot}
\end{equation}
where  $V = 1/\rho$ is the specific volume. The choice of the chemical
potential is natural if we use $p$ and $T$  as variables. In fact from eqs.
(\ref{pot}-\ref{dpot}) it follows
\begin{equation}
V=\mu_{p}\,, \quad S=-\mu_{T}\, ,\quad \varepsilon=\mu-p\mu_{p}-T\mu_{T}\,,
\label{variabili}
\end{equation}
where
\[
\mu_{p}= \left(\frac{\partial \mu}{\partial p}\right)_{T},\ \quad \mu_{T}=
\left(\frac{\partial \mu}{\partial T} \right)_{p}\, .
\]
From eqs. $(\ref{variabili})_{1}-(\ref{variabili})_{3}$ we get
\begin{equation}
\varepsilon_{p}= - p V_{p}-T V_{T}\,. \label{vare}
\end{equation}
The M\"{u}ller definition of incompressibility (\ref{indp}) and Eq. ({\ref{vare}) imply
  $V$ is constant, i.e.
\[
\rho=\rho_{0}=\textrm{constant}\,.
\]
Obviously, this M\"uller result \cite{Muller1}\, (p. 27) disagrees with the
experiments proving that the volume of fluids changes with the temperature  and also, a little, with the pressure  \cite{Handbook,Fine}.

\section{Removal of the M\"uller paradox for fluids}

According to the fact that the so-called \textit{compressibility coefficient}
$\beta = -V_{p}/V $ is very small in the case of   incompressible body,
to remove the M\"uller paradox we define the notion of quasi
thermal-incompressible fluid:
\newline
\textit{\textbf{A quasi thermal-incompressible fluid}} \textit{is a medium for
which the only   equation independent of $p$ among constitutive equations}
(\ref{const1})-(\ref{const0}) \textit{is the density}:
\begin{equation}
\rho \equiv \rho(T)\, .  \label{density function}
\end{equation}
We assume that the condition of independence of $p$ is not necessary for the
other constitutive equations; in particular, the internal energy $\varepsilon$
remains  function of $p$ and $T$:
\[
\varepsilon= \varepsilon (p,T)\,.
\]
From Eq. (\ref{density function}), by integration of  Eq. $(\ref{variabili})_{1}$, we
obtain
\begin{equation}
\mu (p,T)=  V(T)\ p+\mu_{0}(T)\,, \label{chempot}
\end{equation}
and substituting in Eq. $(\ref{variabili})_{3}$
\begin{equation}
\varepsilon (p,T)= - TV^{\prime}(T)\ p+e(T)\,, \label{intrn1}
\end{equation}
where
\[
 e(T)=\mu_{0}-T \mu_{0}^{\prime}\,,
\]
with\ \, $^{\prime}= d/dT$. Following M\"uller proposal, we name
incompressible fluid  (or better \emph {perfectly incompressible fluid})
 a fluid for which  all the constitutive equations are independent of $p$.
\newline Therefore, from Eq. (\ref{intrn1}), a \emph{quasi
thermal-incompressible fluid}  tends to be
\emph{perfectly incompressible} if $\varepsilon(p,T)$ can be approximated with
$e(T)$, i.e. when:
\begin{equation}
p \ll \frac{e(T)}{|V^{\prime}| T}=\frac{\rho^{2} e(T)}{{| \rho^{\prime} | T}} \,
. \label{relt}
\end{equation}
\newline For example, let us consider the classical linear behavior (also typical  of Boussinesq approximation):
\begin{equation}
V = V_{0}[1+\alpha (T-T_0)] ,  \label{ff}
\end{equation}
where $\alpha$ is the \textit{thermal expansion}, the constant $V_0=1/\rho_{0}$ is associated with the scale of volume and $T_0$ is a reference temperature.
If we assume that the  specific heat at constant pressure  $c_p \equiv h_T$ is constant   (where $h= \varepsilon + p V$ is the enthalpy), then from Eq. (\ref{intrn1}) we have
\begin{equation}
e(T)= c_p\, T. \label{cpnocv}
\end{equation}
Inserting eqs. (\ref{ff}-\ref{cpnocv}) into inequality (\ref{relt}) we obtain:
\begin{equation}
p \ll p_{cr}\quad \mathrm{with} \quad p_{cr}= {c_{p}\, \rho_{0}}/{\alpha} \, .
\label{crit}
\end{equation}
We call $p_{cr}$ the critical pressure at density $\rho_{0}$. As we expect in physical situations, we note that $p_{cr}$
is inversely proportional to $\alpha$.
Critical pressure $p_{cr}$ and inequality (\ref{crit})  characterize the fact that  a quasi thermal-incompressible fluid is experimentally similar to a
perfectly incompressible fluid. In such a case, the M\"uller paradox is
removed.

\subsection{Application to water}

To evaluate the magnitude order of the critical pressure (\ref{crit}), we give
some numerical results in the case of water.\newline At temperature $T_0=
20 {^\circ}$ C, we get \cite{Handbook,Handbook2}:
\[
\rho_{0} \simeq 10^{3} \mathrm{kg/m}^{3},\quad c_{p} \simeq 4.2 \cdot 10^{3}
\mathrm{Joule/kg.K},\quad \alpha= 207 \cdot 10^{-6} \mathrm{/K},
\]
and from Eq. (\ref{crit}) we deduce: $p_{cr} \simeq 2 \cdot 10^{10}$ Pascal
$\simeq 2 \cdot 10^{5}$ atm.
\newline
The value of critical pressure is  large  with respect to the normal pressure
conditions. The fact that for usual pressures, a liquid is experimentally incompressible and
the volume changes  with the  temperature  ( Eq. (\ref{ff}) ) does
not violate the principles of thermodynamics.
We can interpret the result in another way: for very large pressure, perfectly incompressible fluids do not exist. From  physical point of view, this observation seems
  reasonable,  for example in astrophysics where very high pressures are
present.
\newline
We also observe that from    experimental data of sound velocity in the water  \cite{Fine,Handbook2},  the compressibility coefficient $\beta$  is not zero but very small  ($\beta =  4.98\times 10^{-10}/\,\texttt{Pascal}$). This is in agreement with the fact that perfect incompressibility is  an idealization.
%\newline
\vspace{.3cm}

\parindent 0pt
{\textbf{Remark}} \label{remark}

Our definition of quasi thermal-incompressibility is, for
some authors (e.g. \cite{Raj,Raj1}), the definition of incompressible body in the context of Boussinesq approximation.
Nevertheless, we strongly believe that the M\"uller definition of
incompressible fluid is the correct one even if it is a limit case: only in
this case, we obtain the same differential equations as for the usual
incompressible approach with ${\rho,\mathbf{v},T}$ variables. In fact in the case of
incompressibility, i.e. when $\rho\equiv \rho(T)$, a generic
constitutive quantity $\phi\equiv \phi(\rho, T)$  becomes a function only of
$T$.  In our analysis  \textit{perfectly incompressible fluid} is considered as
 limit case of \emph{quasi thermal-incompressible fluid}.

The quasi thermal-incompressibility is obtained as a limit process justifying the compatibility between incompressibility and Gibbs relation when inequality (\ref{crit}) is verified. The quasi thermal-incompressibility does not characterize a real compressible material; for real compressible fluids  the chemical potential $\mu$ must be a concave  function of $(p,T)$. When $V$ depends only on $T$, the chemical potential is a linear function of $p$ (see  Eq. (\ref{chempot}))  and consequently cannot be concave.
For real compressible materials the volume $V$ necessarily depends on $p$; quasi thermal-incompressible materials can be considered as an approximation of incompressible materials when the pressure is sufficiently small such that  inequality (\ref{crit}) is satisfied.
\newline
Moreover, note that we use term of quasi-thermal incompressibility, rather than term of quasi
incompressibility, because this last term has a different meaning in the pure
isothermal mechanical case.

\section{Incompressible hyperelastic media}

In the case of elasticity, M\"uller presented a similar paradox \cite{Muller1}\, (p. 263). In fact, he proved that
\[
J \equiv \rm{det}\, \mathbf{F} ,
\]
where $\mathbf{F}$ denotes the deformation gradient, cannot depend on the
temperature $T$ and must be constant as in the pure mechanical case, i.e. $J=1$.
This result also disagrees with experiments proving that the volume of an
elastic incompressible solid changes with the temperature. \newline We use similar
arguments as in fluid case, but on the contrary of the M\"uller procedure, we
consider a particular decomposition of the stress tensor allowing to remove the
paradox. Moreover, our approach is  also valid for non-isotropic materials.

\subsection{A decomposition of the stress tensor in the case of  hyperelastic medium}

Each particle of the continuous medium is labeled by a material variable {$%
\mathbf{X}$}, ranging from a reference configuration $\mathcal{D}_{0}$ into an
Euclidian space \cite{Wang}. The reference density $\rho _{0}$ is given as a
function on $\mathcal{D}_{0}$ \cite{Germain}. \newline The expression
${\mathbf{x}} =\phi (\mathbf{X},t)$ of the spatial position describes the
motion of the continuous medium. Generally, $\phi (.
,t)$ is a twice continuously differentiable diffeomorphism from $\mathcal{D}%
_{0}$ into a compact oriented manifold $\mathcal{D}_{t}$ constituting the
image of the material at time $t$. As usual, we denote by $\mathbf{C}=%
\mathbf{F}^{T}\mathbf{F}$ the right Cauchy-Green deformation tensor, where
superscript \ $^T$ \ means the transposition. Let us recall:
\begin{equation}
\rho\ (\det \mathbf{C})^{\frac{1}{2}} =\rho _{0}.  \label{mass}
\end{equation}%
For the sake of simplicity, we consider the case of homogeneous bodies (the
case of inhomogeneous bodies can be treated in the same way). The internal
energy density is supposed to be  function of the tensor $\mathbf{C}$ and
temperature $T$
\[
\varepsilon \equiv \varepsilon(\mathbf{C},T),
\]
and the stress tensor can be written \cite{Germain,Ruggeri,Truesdell},
\begin{equation}
{\mathbf{t}} =2\rho \mathbf{F}\frac{\partial \psi}{\partial \mathbf{C}}\,%
\mathbf{F}^{T},  \label{stresstensor}
\end{equation}%
where $\psi=\varepsilon -TS$ is the specific free energy. By writing
\cite{Flory,Gouin}:
\begin{equation}
\mathbf{\widetilde{C}}=\frac{1}{(\det \mathbf{C})^{\frac{1}{3}}}\,\mathbf{C}%
\quad \rm{or} \quad \mathbf{C}= \left(\frac{\rho_0}{\rho}\right)^{\frac{2}{%
3}} \mathbf{\widetilde{C}}\,,  \label{ci}
\end{equation}
the specific free energy   can be expressed in the form:%
\[
\psi \equiv f(\rho ,\mathbf{\widetilde{C}},T).  \label{homogeneneous}
\]%
We note that $\mathbf{C}$  is substituted by the independent variables $\rho $
and $\mathbf{\widetilde{C}}$. Since $\det \mathbf{\widetilde{C}}=1$, the variable
$\rho $ corresponds to the change of volume while the tensorial variable
$\mathbf{\widetilde{C}}$ represents the distortion of the medium. This point is
fundamental for the decomposition of the stress tensor and
will be the key of the demonstration. When $f$ is independent of $\mathbf{\widetilde{C}}$ we are back to the fluid case. It is more convenient to introduce
the function $g$ such that:
\[
g(\rho, \mathbf{C}, T) \equiv f\left(\rho,\frac{1}{(\det \mathbf{C})^{\frac{1%
}{3}}}\, \mathbf{C},T \right).  \label{homogeneneous}
\]
Consequently, $g$ is a homogeneous function of degree zero with respect to $%
\mathbf{C}$.
\newline
From Eqs. (\ref{mass}-\ref{stresstensor}) it follows:%
\begin{equation}
{\mathbf{t}} =2\rho \mathbf{F}\left( \frac{\partial g}{\partial \rho }\frac{%
\partial \rho }{\partial \mathbf{C}}+\frac{\partial g}{\partial \mathbf{C}}%
\right) \mathbf{F}^{T}.  \label{stresstensor2}
\end{equation}
Differentiating Eq. (\ref{mass}) and using Jacobi's identity, we obtain
\[
d\rho =-\frac{1}{2}\rho\, \mathbf{C}^{-1} \cdot d\mathbf{C}\,,
\]
where the dot represents the scalar product between matrices.
Hence,%
\[
\frac{\partial \rho}{\partial \mathbf{C}}=-\frac{1}{2}\rho\, \mathbf{C}^{-1}\,,
\label{Cderivative}
\]%
and consequently, Eq. (\ref{stresstensor2}) yields,
\[
{\mathbf{t}} =-\rho^2 \frac{\partial g}{\partial \rho }\,\mathbf{1}+2\rho
\mathbf{F}\frac{\partial g}{\partial \mathbf{C}}\mathbf{F}^{T}. \label{t}
\]
Due to the fact $g$ is
homogeneous of degree zero with respect to $\mathbf{C}$, from the Euler's identity
we immediately deduce:
\[
\frac{\partial g}{\partial \mathbf{C}} \cdot \mathbf{\ C}=0\,,   \]
and
\begin{equation}
{\mathbf{t}} =-p\mathbf{1}+{\mathbf{\bm{\tau}}}\,,  \label{dcp}
\end{equation}%
with
\begin{equation}
\begin{array}{ccccccc}
\displaystyle p=\rho
%TCIMACRO{\U{b2}}%
%BeginExpansion
{{}^2}%
%EndExpansion
\displaystyle\frac{\partial g}{\partial \rho }, &  & \displaystyle{\mathbf{%
\bm{\tau}}}=2\rho \mathbf{F}\frac{\partial g}{\partial \mathbf{C}}%
\mathbf{F}^{T} &  & \rm{and} &  & \rm {tr}\, {\mathbf{\bm{\tau}}}=0,%
\end{array}
\label{sigma}
\end{equation}
where $\textsf{tr}$ is the trace operator.
Let us note that, in relation (\ref{dcp}), ${\mathbf{t}}$ is similar to the fluid decomposition of Eq. (\ref{fluidtensor}) and in the
solid case, $p$ is analog to a pressure.

 The
decomposition (\ref{dcp}-\ref{sigma}) allows  to define
a pressure also in the case of an elastic body.

\subsection{Removal of the M\"uller paradox for hyperelastic media}

The Gibbs equation in the case of elastic materials is
\cite{Germain,Ruggeri,Truesdell}
\begin{equation}
TdS=d\varepsilon -\frac{1}{2\rho_{0} }\mathbf{S} \cdot d\mathbf{C} \,
,\label{gel}
\end{equation}%
where
\begin{equation}
\textbf{S}= J\textbf{F}^{-1}\textbf{t}\,(\textbf{F}^{T})^{-1}  \label{esse}
\end{equation}
is the second Piola-Kirchhoff stress tensor. Inserting Eq. ({\ref{dcp}) into Eq.
({\ref{esse}), we get from Eq. ({\ref{gel}) the Gibbs relation:
\begin{equation}
TdS=d\varepsilon -\frac{p}{\rho ^{2}}\,d\rho -\frac{1}{2\,\rho }
 \mathbf{F}^{-1}{\mathbf{\bm{\tau }}}\mathbf{F}^{-1^{T}} \cdot d\mathbf{C} .  \label{Gibbs 1}
\end{equation}%
From Eq. (\ref{ci}) we obtain:
\[
d\mathbf{C=-}\frac{2}{3}\frac{\mathbf{C}}{\rho }\,d\rho +\left( \frac{\rho
_{0}}{\rho }\right) ^{\frac{2}{3}}d\mathbf{\widetilde{C}},
\]%
and Eq. (\ref{Gibbs 1}) leads to:
\[
TdS=d\varepsilon -\frac{p}{\rho ^{2}}\,d\rho +\frac{1}{3\,\rho ^{2}}
\mathbf{F}^{-1}{\mathbf{\bm{\tau }}}\, \mathbf{F}^{-1^{T}}\cdot
\mathbf{C}\  d\rho -\frac{1}{2\,\rho } \left( \frac{%
\rho _{0}}{\rho }\right) ^{\frac{2}{3}}\mathbf{F}^{-1}{\mathbf{\bm{%
\tau }}}\,\mathbf{F}^{-1^{T}}\cdot d\mathbf{\widetilde{C}} .
\]%
From Eq. (\ref{sigma}), we get\ $\mathbf{F}^{-1}{\mathbf{\bm{\tau
}}}\mathbf{F}^{-1^{T}}
\cdot \mathbf{C}= \mathtt{tr}\left( \mathbf{F}^{-1}{\mathbf{%
\bm{\tau }}}\mathbf{F}^{-1^{T}}\mathbf{C} \right)\equiv \mathtt{tr}%
 \, { \mathbf{\bm{\tau
 }} } =0.$ If we define
\begin{equation}
{\mathbf{\bm{\widetilde{\tau }}}}=-\frac{1}{2\,\rho }\left( \frac{\rho _{0}}{%
\rho }\right) ^{\frac{2}{3}}\mathbf{F}^{-1}{\mathbf{\bm{\tau }}}%
\mathbf{F}^{-1^{T}}, \label{sigmap}
\end{equation}
we obtain the Gibbs relation in the final form:
\begin{equation}
TdS=d\varepsilon -\frac{p}{\rho ^{2}}\,d\rho +{\mathbf{\bm{\widetilde{\tau}
}}}\cdot d\mathbf{\widetilde{C}} .  \label{Gibbs2}
\end{equation}
As in the case of fluid, introducing the chemical potential
\[
\mu= \varepsilon+pV+\bm{\widetilde{\tau}} \cdot \mathbf{\widetilde{C}}-TS\, ,
\]
Eq. (\ref{Gibbs2}) implies
\begin{equation}
d\mu= V dp-SdT +\mathbf{\widetilde{C}}\cdot d\bm{\widetilde{\tau}}.
\label{difmu}
\end{equation}
Equation  (\ref{sigmap}) together with Eq. $(\ref{sigma})_{3}$, imply
$
\textsf{tr}(\bm{\widetilde{\tau}}\, \textbf{C})=0\,.
$
Therefore $\bm{\widetilde{\tau}}$ has only five independent components.
\newline
From Eq. (\ref {difmu}) it is natural to introduce the change of variables ($\rho, T,
\mathbf{\widetilde{C}}$) into ($p, T, \bm{\widetilde{\tau}}$). In fact, Eq. (\ref
{difmu}) implies:
\begin{equation}
 V= \mu_{p}\,, \quad \mathbf{\widetilde{C}}= \mu_{\bm{\widetilde{\tau}}}\,, \quad S=-\mu_{T}\,, \quad
\varepsilon = \mu-T \mu_{T}-p\,\mu_{p}-\bm{\widetilde{\tau}}\cdot \mu
_{\bm{\widetilde{\tau}}}\, , \label{varie}
\end{equation}
where now
\[
\mu _{p} =\left( \frac{\partial \mu}{\partial p}\right)
_{T,\,\bm{\widetilde{\tau}}},\quad  \mu _{\bm{\widetilde{\tau}}}=\left(
\frac{\partial \mu }{\partial \bm{\widetilde{\tau}}}\right) _{p,\,T},\quad
\mu _{T}=\left( \frac{\partial \mu}{\partial T}\right)
_{p,\,\bm{\widetilde{\tau}}}\, .
\]
By analogy with the case of fluids, \emph{\textbf{a quasi
thermal-incompressible elastic medium}} \emph{must verify  the condition}}:
\[
J\equiv J(T)\quad  \emph{or\ \, equivalently} \quad V\equiv V (T)\,,
\]%
\emph{while the other constitutive equations still depend on}
$(p,T,\bm{\widetilde{\tau}})$.
In particular
\[
\varepsilon \equiv \varepsilon (p,T,\bm{\widetilde{\tau}})\,.
\]%
By integration of Eq. ($\ref{varie})_{1}$, we obtain:
\[
 \mu (p, T, \bm{\widetilde{\tau}})= V(T)\,p+\mu_{0}(T,
 \bm{\widetilde{\tau}})\, ,
\]
and by substituting in Eq. ($\ref{varie})_{4}$, we get
\begin{equation}
 \varepsilon= -T V^{\prime}(T)\,p+ e(T, \bm{\widetilde{\tau}})\, , \label{intr}
\end{equation}
with
\[
e(T, \bm{\widetilde{\tau}})=\mu_{0}-T
\mu_{{0T}}-\bm{\widetilde{\tau}}\cdot
\mu_{{0}{\bm{\widetilde{\tau}}}}\,.
\]
Following M\"uller proposal again, we define an \textit{incompressible elastic
medium} (or better \textit{perfectly incompressible elastic medium}) as a solid
for which all the constitutive equations are independent of $p$.
\newline
Therefore, taking account of Eq. (\ref{intr}), a thermal-incompressible elastic
medium tends to be \emph{perfectly incompressible} if $\varepsilon
(p,T,\bm{\widetilde{\tau}})$ can be approximated by
$e(T,\bm{\widetilde{\tau}})$. This assumption is verified when
\[
p\ll \frac{e(T,\bm{\widetilde{\tau}})}{\displaystyle |\, V^{\prime}| {\
T}}\,.
\]%
We consider the case:
\[
e(T,\bm{\widetilde{\tau}})=c_{p}\, T\,\,,\quad V(T)=V_{0}[1+\alpha (T-T_0)]\,,
\]%
where  now,  $\alpha $ and $V_{0}$ are positive constants and $c_{p}$  can be a
function of $\bm{\widetilde{\tau}}$. As well as for  fluids, we obtain:
\begin{equation}
p\ll {c_{p}\rho _{0}}/{\alpha }.  \label{crit21}
\end{equation}%
We assume that $c_{p}\, \rho _{0}/{\alpha}$, varying with $%
\bm{\widetilde{\tau}}$, has a minimal value $p_{cr}$; we   again denote by
$p_{cr}$ the critical pressure. In this case the critical pressure $p_{cr}$ and
inequality $p\ll p_{cr} $ are   characteristic of  quasi thermal-incompressible  elastic
medium to be experimentally similar to   perfectly incompressible one and  the M\"{u}ller paradox is removed.

\subsection{Application to pure gum rubber}

The most famous incompressible hyperelastic medium is the pure gum rubber.\
This material was studied by many authors and in particular by P.J. Flory,
Nobel prize of Chemistry in 1974 \cite{Flory2}.
\newline
In the range of temperatures $[50^{\circ} \,\mathrm{{C},85^{\circ}\,{C}]}$
physical constants are \cite{Handbook2}:
\[
\rho_{0} \simeq 930\, \mathrm{kg/m}^{3},\quad c_{p} \simeq 1.9 \cdot 10^{3}
\mathrm{Joule/kg.K},\quad \alpha= 6.7 \cdot 10^{-3} \mathrm{/K} .
\]
Equation (\ref{crit21}) allows to obtain: $p_{cr}\simeq 2.7\cdot 10^{8}$ Pascal
$\simeq2.7\cdot 10^{3}$  atm.
The critical pressure is very large and therefore pure gum rubber is a good incompressible body in normal conditions with usual pressures.

\section{Conclusions}

In the case of fluids and elastic media, we showed that,  for
pressures smaller than a critical value, the  volume can depend on the temperature
if incompressibility  is defined as limit case of quasi
thermal-incompressibility.
Quasi thermal-incompressible materials can be considered as an approximation of incompressible materials - in the sense of \cite{Raj,Raj1} where $\beta = 0$ - if the pressure is  small enough such that  inequality (\ref{crit}) is satisfied.
\newline
To obtain these results, we used   temperature as
natural thermodynamical variable. Nevertheless, Manacorda \cite{Manacorda} first noted (see also \cite{Scott1,Scott2}) that in the case $V  \equiv V(T)$,  instabilities occur in wave propagations. The instabilities  are due to the chemical potential non-concavity
(see Remark in previous section) and the sound velocity $c$ becomes complex. For this reason, some authors consider the volume as  function of entropy $V \equiv V(S)$ instead of  function of temperature   \cite{Chadwick,Scott}. This assumption does not seem realistic:   $V=V(S)$ cannot be measured because entropy is not an observable  and moreover, in this case $1/c=0$ (i.e., the sound velocity  is infinite); as a consequence  the mathematical structure of Euler fluids becomes  parabolic.

Our goal was to present the simplest model for removing the thermodynamical paradox when the volume depends only on $T$. Nevertheless,
a more realistic definition of quasi
thermal-incompressibility needs to suppose that the compressibility coefficient $\beta$ is small but not zero; the concavity of the chemical potential can be restored and the sound velocity can be real with consequence that  incompressible body can be seen as a limit  case of a compressible one but the present result cannot quantitatively change. This will be the subject of a forthcoming study.
\newline
Finally  thanks to decomposition (\ref{dcp}),  the technique used to remove the
M\"uller paradox is available both for  fluids and elastic media. This similarity
allows to forecast a possible Boussinesq approximation  in case of elastic
media.

\bigskip

{\small{\textbf{Acknowledgments}:
The authors are grateful to Professor Salvatore Rionero for his interest on this paper
and his useful reference suggestions.}}

\end{document}